\documentclass[aps,preprint,nofootinbib]{revtex4}%
\usepackage{amsfonts}
\usepackage{amsmath}
\usepackage{amssymb}
\usepackage{graphicx}%
\providecommand{\U}[1]{\protect\rule{.1in}{.1in}}

\begin{document}
\preprint{ }
\title[Short title for running header]{On canonical transformations between equivalent Hamiltonian formulations of
General Relativity}
\author{A.M. Frolov}
\email{afrolov@uwo.ca}
\affiliation{Department of Chemistry, University of Western Ontario, N6A 5B7, London,
Canada }
\author{N. Kiriushcheva}
\email{nkiriush@uwo.ca}
\affiliation{Faculty of Arts and Social Science, Huron University College, N6G 1H3 and
Department of Applied Mathematics, University of Western Ontario, N6A 5B7,
London, Canada}
\author{S.V. Kuzmin}
\email{skuzmin@uwo.ca}
\affiliation{Faculty of Arts and Social Science, Huron University College, N6G 1H3 and
Department of Applied Mathematics, University of Western Ontario, N6A 5B7,
London, Canada}
\pacs{PACS number}

\begin{abstract}
Two Hamiltonian formulations of General Relativity, due to Pirani, Schild and
Skinner (Phys. Rev. 87, 452, 1952) and Dirac (Proc. Roy. Soc. A 246, 333,
1958), are considered. Both formulations, despite having different expressions
for constraints, allow one to derive four-dimensional diffeomorphism
invariance. The relation between these two formulations at all stages of the
Dirac approach to the constrained Hamiltonian systems is analyzed. It is shown
that the complete sets of their phase-space variables are related by a
transformation which satisfies the ordinary condition of canonicity known for
unconstrained Hamiltonians and, in addition, converts one total Hamiltonian
into another, thus preserving form-invariance of generalized Hamiltonian
equations for constrained systems.

\end{abstract}
\eid{identifier}
\date{\today}
\maketitle


\section{Introduction}

The Hamiltonian formulation of General Relativity (GR) is an old subject which
is still plagued by some long-standing questions. One of the most important
problems, related to essence of Einstein's General Relativity, was the
disappearance of four-dimensional diffeomorphism%

\begin{equation}
\delta g_{\mu\nu}=-\xi_{\mu;\nu}-\xi_{\nu;\mu} \label{eqnE0}%
\end{equation}
in the Hamiltonian formulation of GR \textquotedblleft that has worried many
people working in geometrodynamics for so long\textquotedblright%
\ \cite{Isham}\footnote{The word \textquotedblleft
diffeomorphism\textquotedblright\ is often used as equivalent to the
transformation (\ref{eqnE0}) (the semicolon \textquotedblleft$;$%
\textquotedblright\ means a covariant derivative); in this paper the same
meaning is employed.}. According to some authors, in the Hamiltonian
formulation of GR, it is possible to restore only spatial diffeomorphism
\cite{DeWitt, Pullin} or, according to others, the so-called \textquotedblleft
special diffeomorphism\textquotedblright, for which a non-covariant and
field-dependent redefinition of gauge parameters is needed, can be derived
\cite{Bergmann, Pons, Saha}.

In fact, the two Hamiltonian formulations that preserve four-dimensional
diffeomorphism have been known for a long time. They are the first Hamiltonian
formulations of GR due to Pirani, Schild, and Skinner (PSS) \cite{Pirani} and
Dirac \cite{Dirac}, both of which allow one to derive the full diffeomorphism
from their constraint structure \cite{KKRV, myths}. These two formulations
both lead to the expected gauge invariance (\ref{eqnE0}). At the same time,
they provide an example that allows us to discuss the conditions under which
different Hamiltonian formulations of GR are equivalent. The study of the
conditions for which a change of phase-space variables preserves the
properties of an original Hamiltonian system is of great importance for
constrained dynamical systems. It is especially important in the Hamiltonian
formulations of General Relativity where it is customary to perform changes of
variables or to introduce new variables. The legitimacy of such changes must
be verified. Without showing that a change of variables is canonical, it is
impossible to assert that the Hamiltonian formulation of a new theory is
equivalent to the original one. This what happened with the formulation that
prevailed all others for more than fifty years: the ADM gravity \cite{ADM}.
The introduction of \textquotedblleft lapse\textquotedblright\ and
\textquotedblleft shift\textquotedblright\ functions: $N=\left(
-g^{00}\right)  ^{-1/2}$ and $N^{i}=-\frac{g^{0i}}{g^{00}}$, which is in fact
nonlinear and non-covariant transformation, leads to the formulation (ADM)
that is not related to Dirac's (and consequently, to PSS) formulation by any
canonical transformation. As we demonstrated in \cite{myths}, one Poisson
bracket is enough to prove the non-canonicity of the ADM variables, e.g.%

\begin{equation}
\left\{  N\left(  x\right)  ,\Pi^{ki}\left(  x^{\prime}\right)  \right\}
=\left\{  \left(  -g^{00}\right)  ^{-1/2},p^{ki}\right\}  _{g,p}=-\frac
{\delta}{\delta g_{ki}}\left(  -g^{00}\right)  ^{-1/2}=\frac{1}{2}\left(
-g^{00}\right)  ^{-3/2}g^{0k}g^{0i}\neq0 \label{eqn445}%
\end{equation}
(where $\Pi^{ki}$ and $p^{ki}$ are momenta in ADM and Dirac's formulations,
respectively, conjugate to the space-space components of the metric tensor,
$g_{ki}$).

As the result, the ADM formulation (the only one in which a restoration of
gauge invariance from the complete set of first class constraints has even
been considered before) does not produce the expected diffeomorphism
invariance (\ref{eqnE0}), as it was recently demonstrated in \cite{Saha} using
the method \cite{Novel}. The derivation of \cite{Saha} is the most complete
one in the literature but it is not new; the gauge transformations of the ADM
variables have been discussed in part previously in \cite{Bergmann} and
\cite{Castellani}. The gauge transformations that follow from the ADM
formulation can be presented \textit{in the form} of diffeomorphism only if a
\textit{field-dependent} and \textit{non-covariant} redefinitions of gauge
parameters are performed%
\begin{equation}
\xi_{diff}^{0}=\left(  -g^{00}\right)  ^{1/2}\varepsilon_{ADM}^{\perp
}~,\left.  {}\right.  \left.  {}\right.  \xi_{diff}^{k}=\varepsilon_{ADM}%
^{k}+\frac{g^{0k}}{g^{00}}\left(  -g^{00}\right)  ^{1/2}\varepsilon
_{ADM}^{\bot}~, \label{redef}%
\end{equation}
which, according to \cite{Saha}, \textquotedblleft demonstrate the unity of
the different symmetries involved\textquotedblright. The transformations of
\cite{Saha} are consistent with transformations obtained in \cite{Banerjee}
using the Lagrangian approach of \cite{Gitman}. However, as we have already
pointed out in \cite{affine-metric}, the field-dependent redefinition of gauge
parameters contradicts the essence of all known algorithms of restoration of
gauge invariance, as all of them start from the assumption that the gauge
parameters should be independent of fields (for details see
\cite{affine-metric}). To prevent the situation where some manipulations are
needed to justify non-canonical transformations, it is better not to perform
such transformations from the beginning.

To the best of our knowledge, the equivalence of Hamiltonian formulations of
GR, which differ from each other by a change of phase-space variables, was
never been analyzed. What is only known to us is a brief statement of
DeWitt,\ which he made for PSS formulation: \textquotedblleft four so-called
primary constraints could, by a phase transformations, be changed into pure
momenta\textquotedblright\ (see \cite{DeWitt} where the author refers to his
unpublished report). The connection between the linearized versions of the two
formulations of \cite{Pirani} and \cite{Dirac} was analyzed in \cite{GKK}
where it was demonstrated that the two formulations of linearized GR are
connected by a change of phase-space variables, which is in fact a canonical
transformation in the sense of ordinary Classical Mechanics. Moreover, these
formulations, despite having different expressions for Hamiltonians and
constraints, give an equivalent description, i.e. the corresponding generators
built from the first-class constraints allow one to derive the same gauge
invariance (the linearized version of diffeomorphism invariance).

The main goal of this paper is to extend this analysis to the two Hamiltonian
formulations of full GR \cite{Pirani, KKRV, Dirac, myths}. We investigate the
relation between the corresponding phase-space variables in both formulations
and discuss the effects of such a change of variables at all stages of the
Dirac procedure. Another aim is to formulate some general conditions that
should be imposed on transformations of phase-space variables for singular
systems to preserve the equivalence of different Hamiltonian formulations.

\section{Comparison of the two Hamiltonian formulations of GR}

A starting point of the Hamiltonian formulations of GR in both the approaches
of \cite{Pirani} and \cite{Dirac} is the \textquotedblleft
gamma-gamma\textquotedblright\ part of the Einstein-Hilbert (EH) Lagrangian
which is quadratic in first-order derivatives of the metric tensor (for more
details see, e.g., \cite{Landau, Carmeli})
\begin{equation}
L=\sqrt{-g}g^{\alpha\beta}\left(  \Gamma_{\alpha\nu}^{\mu}\Gamma_{\beta\mu
}^{\nu}-\Gamma_{\alpha\beta}^{\nu}\Gamma_{\nu\mu}^{\mu}\right)  =\frac{1}%
{4}\sqrt{-g}B^{\alpha\beta\gamma\mu\nu\rho}g_{\alpha\beta,\gamma}g_{\mu
\nu,\rho} \label{eqn1}%
\end{equation}
where
\begin{equation}
B^{\alpha\beta\gamma\mu\nu\rho}=g^{\alpha\beta}g^{\gamma\rho}g^{\mu\nu
}-g^{\alpha\mu}g^{\beta\nu}g^{\gamma\rho}+2g^{\alpha\rho}g^{\beta\nu}%
g^{\gamma\mu}-2g^{\alpha\beta}g^{\gamma\mu}g^{\nu\rho}. \label{eqn2}%
\end{equation}
To find the momenta $\pi^{\alpha\beta}$, conjugate to the ten components of
the metric tensor $g_{\alpha\beta}$, we rewrite Eq. (\ref{eqn1}) in a form
which explicitly contains the time derivatives of the metric tensor, i.e. in
terms of \textquotedblleft velocities\textquotedblright\
\begin{equation}
L=\frac{1}{4}\sqrt{-g}B^{\alpha\beta0\mu\nu0}g_{\alpha\beta,0}g_{\mu\nu
,0}+\frac{1}{2}\sqrt{-g}B^{\left(  \alpha\beta0\mid\mu\nu k\right)  }%
g_{\alpha\beta,0}g_{\mu\nu,k}+\frac{1}{4}\sqrt{-g}B^{\alpha\beta k\mu\nu
t}g_{\alpha\beta,k}g_{\mu\nu,t}~, \label{eqn2.1}%
\end{equation}
where the Latin alphabet is used for spatial components and \textquotedblleft%
0\textquotedblright\ for a temporal one. The brackets $(\alpha\beta)$ indicate
symmetrization in two indices, while the notation $\left(  ...\mid...\right)
$ is used for symmetrization in two groups of indices, i.e.
\[
B^{\left(  \alpha\beta\gamma\mid\mu\nu\rho\right)  }=\frac{1}{2}\left(
B^{\alpha\beta\gamma\mu\nu\rho}+B^{\mu\nu\rho\alpha\beta\gamma}\right)  .
\]
Momenta conjugate to the metric tensor are defined in standard way, and
(\ref{eqn2.1}) gives
\begin{equation}
\pi^{\gamma\sigma}=\frac{\delta L}{\delta g_{\gamma\sigma,0}}=\frac{1}{2}%
\sqrt{-g}B^{\left(  \left(  \gamma\sigma\right)  0\mid\mu\nu0\right)  }%
g_{\mu\nu,0}+\frac{1}{2}\sqrt{-g}B^{\left(  \left(  \gamma\sigma\right)
0\mid\mu\nu k\right)  }g_{\mu\nu,k}~. \label{eqn3}%
\end{equation}
By using (\ref{eqn2}), one finds the explicit form of the first term of
(\ref{eqn3})
\begin{equation}
B^{\left(  \left(  \gamma\sigma\right)  0\mid\mu\nu0\right)  }g_{\mu\nu
,0}=g^{00}E^{\mu\nu\gamma\sigma}g_{\mu\nu,0} \label{eqn3.1}%
\end{equation}
where
\begin{equation}
E^{\mu\nu\gamma\sigma}\equiv e^{\mu\nu}e^{\gamma\sigma}-e^{\mu\gamma}%
e^{\nu\sigma},\left.  {}\right.  e^{\mu\nu}\equiv g^{\mu\nu}-\frac{g^{0\mu
}g^{0\nu}}{g^{00}}. \label{eqn3.2}%
\end{equation}
Note that both $e^{\mu\nu}$ and $E^{\mu\nu\gamma\sigma}$ are zero unless all
of the $\mu$, $\nu$, $\gamma$, and $\sigma$ indices differ from $0$. The
notation $e^{km}$ designates the inverse of the spatial components of the
metric tensor, i.e. $g_{nk}e^{km}=\delta_{n}^{m}$, and $\frac{\delta}{\delta
g_{0\alpha}}e^{\mu\nu}=\frac{\delta}{\delta g_{0\alpha}}E^{\mu\nu\gamma\sigma
}=0$. From (\ref{eqn3.1})-(\ref{eqn3.2}) it follows that we cannot express
some of the velocities in (\ref{eqn3}) in terms of momenta, therefore, $d$
primary constraints arise (here $d$ is the dimension of space-time); they are
\begin{equation}
\phi^{0\sigma}=\pi^{0\sigma}-\frac{1}{2}\sqrt{-g}B^{\left(  \left(
0\sigma\right)  0\mid\mu\nu k\right)  }g_{\mu\nu,k}\approx0. \label{eqn4}%
\end{equation}
\qquad\qquad\qquad

If $\gamma$ and $\delta$ indices in (\ref{eqn3}) are space-like, then
(\ref{eqn3}) is invertible and we find
\begin{equation}
g_{mn,0}=I_{mnpq}\frac{1}{g^{00}}\left(  \frac{2}{\sqrt{-g}}\pi^{pq}%
-B^{\left(  \left(  pq\right)  0\mid\mu\nu k\right)  }g_{\mu\nu,k}\right)
\label{eqn5}%
\end{equation}
where
\begin{equation}
I_{mnpq}\equiv\frac{1}{d-2}g_{mn}g_{pq}-g_{mp}g_{nq},\left.  {}\right.
I_{mnpq}E^{pqkl}=\delta_{m}^{k}\delta_{n}^{l}~. \label{eqn6}%
\end{equation}
The appearance of a singularity in (\ref{eqn6}) for $d=2$ corresponds to the
fact that in two dimensions none of the components of (\ref{eqn3}) can be
solved for \textquotedblleft velocities\textquotedblright. The number of
primary constraints (three) in this case equals the number of independent
components of the metric tensor in two dimensions \cite{OnHam, 2D}.

The Hamiltonian is defined by $H=\pi^{\alpha\beta}g_{\alpha\beta,0}-L$. After
using (\ref{eqn5}) to eliminate all the \textquotedblleft
velocities\textquotedblright\ $g_{mn,0}$, one finds the following total
Hamiltonian:
\begin{equation}
H_{T}=H_{c}+g_{00,0}\phi^{00}+2g_{0k,0}\phi^{0k}, \label{eqn6-2}%
\end{equation}
where the `canonical part'\footnote{We shall use this standard terminology, or
alternatively `canonical Hamiltonian', both of which however, can be
misleading because for the canonical treatment of singular systems the total
Hamiltonian, $H_{T}$, not its individual parts, is needed to provide the
complete description.} is
\[
H_{c}=\frac{1}{\sqrt{-g}g^{00}}I_{mnpq}\pi^{mn}\pi^{pq}-\frac{1}{g^{00}%
}I_{mnpq}\pi^{mn}B^{\left(  pq0\mid\mu\nu k\right)  }g_{\mu\nu,k}%
\]%
\begin{equation}
+\frac{1}{4}\sqrt{-g}\left[  \frac{1}{g^{00}}I_{mnpq}B^{\left(  \left(
mn\right)  0\mid\mu\nu k\right)  }B^{\left(  pq0\mid\alpha\beta t\right)
}-B^{\mu\nu k\alpha\beta t}\right]  g_{\mu\nu,k}g_{\alpha\beta,t}~.
\label{eqn7}%
\end{equation}
For the detailed analysis of (\ref{eqn6-2}), including the constraint
structure and derivation of the corresponding generators and gauge
transformations, see \cite{KKRV}.

In this Letter we want to compare the Hamiltonian formulation of GR given by
(\ref{eqn6-2}) with that of Dirac \cite{Dirac, myths}. Dirac's main idea was
based on the fact that the Lagrangian, (\ref{eqn1}) (it is called below
$L_{PSS}$) can be modified in order to simplify the primary constraints by
adding a non-covariant combination of spatial and temporal derivatives that
does not affect the equations of motion. This modification leads to the
following Lagrangian
\begin{equation}
L_{D}=L_{PSS}-L^{\ast} \label{eqnE8}%
\end{equation}
where $L^{\ast}$ \cite{Dirac} is taken by Dirac to be
\begin{equation}
L^{\ast}=\left[  \left(  \sqrt{-g}g^{00}\right)  _{,k}\frac{g^{0k}}{g^{00}%
}\right]  _{,0}-\left[  \left(  \sqrt{-g}g^{00}\right)  _{,0}\frac{g^{0k}%
}{g^{00}}\right]  _{,k}. \label{eqnE9}%
\end{equation}

The explicit form of (\ref{eqnE9}) can be found using the identity
$F_{,\gamma}=\frac{\delta F}{\delta g_{\mu\nu}}g_{\mu\nu,\gamma}$ for the
metric-dependent functional and rewriting the contravariant components of the
metric tensor in terms of $e^{\alpha\beta}$ (see (\ref{eqn3.2})). Finally, we
find
\begin{equation}
L^{\ast}=\frac{1}{2}\sqrt{-g}A^{\alpha\beta0\mu\nu k}g_{\alpha\beta,0}%
g_{\mu\nu,k}~, \label{eqnE10}%
\end{equation}
where we have introduced the following notation
\begin{equation}
A^{\alpha\beta0\mu\nu k}=e^{\alpha\beta}e^{k\mu}g^{0\nu}-e^{\mu\nu}e^{k\alpha
}g^{0\beta}+e^{k\alpha}\frac{g^{0\mu}g^{0\nu}g^{0\beta}}{g^{00}}-e^{k\mu}%
\frac{g^{0\alpha}g^{0\nu}g^{0\beta}}{g^{00}}. \label{eqnE11}%
\end{equation}
This relation is obtained by taking into account symmetries $\alpha
\beta\Leftrightarrow\beta\alpha$ and $\mu\nu\Leftrightarrow\nu\mu$ in
(\ref{eqnE10}) due to the presence of $g_{\alpha\beta,0}g_{\mu\nu,k}$. The
important property of $A^{\alpha\beta0\mu\nu k}$ is its antisymmetry with
respect to interchange of the two pairs of indices
\begin{equation}
A^{\alpha\beta0\mu\nu k}=-A^{\mu\nu0\alpha\beta k}. \label{eqnE12}%
\end{equation}
Using the explicit form of (\ref{eqn2}) we can rewrite the coefficients
$B^{\left(  \alpha\beta0\mid\mu\nu k\right)  }$ in terms of the $A^{\alpha
\beta0\mu\nu k}$ and $E^{\alpha\beta\mu\nu}$
\begin{equation}
B^{\left(  \alpha\beta0\mid\mu\nu k\right)  }=A^{\alpha\beta0\mu\nu k}%
+g^{0k}E^{\alpha\beta\mu\nu}-2g^{0\mu}E^{\alpha\beta k\nu}. \label{eqnE13}%
\end{equation}

Now, the relation between Dirac's Lagrangian $L_{D}$ and the Lagrangian of
PSS, $L_{PSS}$, takes the form
\begin{equation}
L_{D}=L_{PSS}-\frac{1}{2}\sqrt{-g}A^{\alpha\beta0\mu\nu k}g_{\alpha\beta
,0}g_{\mu\nu,k}~. \label{eqnE14}%
\end{equation}
Note that at the Lagrangian level the difference between the PSS and Dirac
formulations does not affect the equations of motion and, in this sense, the
two formulations are equivalent. Now, we analyze this difference from the
point of view of the Hamiltonian formulation. If we have the two Lagrangians,
then we can introduce the two corresponding Hamiltonians which, as we know,
give the same gauge invariance \cite{KKRV, myths}. Let us find the relation
between their phase-space variables and constraints. This will provide a clue
about the changes which can be performed at the Hamiltonian level in a
constrained system that will preserve its properties.

The two Lagrangians in (\ref{eqnE14}) differ from each other by the terms
linear in time derivatives of the metric tensor; this will affect the
expression for conjugate momenta in these two Hamiltonian formulations. For
PSS we have
\begin{equation}
\pi^{\gamma\sigma}=\frac{\delta L_{PSS}}{\delta g_{\gamma\sigma,0}},
\label{eqnE15}%
\end{equation}
while for the Dirac formulation the momentum is
\begin{equation}
p^{\gamma\sigma}=\frac{\delta L_{D}}{\delta g_{\gamma\sigma,0}}=\frac{\delta
L_{PSS}}{\delta g_{\gamma\sigma,0}}+\frac{\delta L^{\ast}}{\delta
g_{\gamma\sigma,0}}. \label{eqnE16}%
\end{equation}
To obtain the relation between these two momenta, we subtract the last two
equations, which gives
\begin{equation}
\pi^{\gamma\sigma}-p^{\gamma\sigma}=\frac{\delta}{\delta g_{\gamma\sigma,0}%
}\left(  \frac{1}{2}\sqrt{-g}A^{\alpha\beta0\mu\nu k}g_{\alpha\beta,0}%
g_{\mu\nu,k}\right)  \label{eqnE17}%
\end{equation}
or
\begin{equation}
p^{\gamma\sigma}=\pi^{\gamma\sigma}-\frac{1}{2}\sqrt{-g}A^{\left(
\gamma\sigma\right)  0\mu\nu k}g_{\mu\nu,k}~. \label{eqnE18}%
\end{equation}
Equation (\ref{eqnE18}) represents the transformation of phase-space variables
for two Hamiltonian formulations of GR, \cite{Pirani} and \cite{Dirac}.

Thus, we have two Hamiltonians with two sets of phase-space variables,
$\left(  g_{\alpha\beta},\pi^{\alpha\beta}\right)  $ and $\left(
g_{\alpha\beta},p^{\alpha\beta}\right)  $; the momenta of these two sets are
connected by the transformation of (\ref{eqnE18}) and the components of the
metric tensor are identical in both formulations. The two sets of fundamental
Poisson brackets (PB) are:
\begin{equation}
\left\{  g_{\alpha\beta},\pi^{\mu\nu}\right\}  =\frac{1}{2}\left(
\delta_{\alpha}^{\mu}\delta_{\beta}^{\nu}+\delta_{\alpha}^{\nu}\delta_{\beta
}^{\mu}\right)  \equiv\Delta_{\alpha\beta}^{\mu\nu},\left.  {}\right.
\left\{  g_{\alpha\beta},g_{\mu\nu}\right\}  =\left\{  \pi^{\alpha\beta}%
,\pi^{\mu\nu}\right\}  =0 \label{eqnE19}%
\end{equation}
and
\begin{equation}
\left\{  g_{\alpha\beta},p^{\mu\nu}\right\}  =\Delta_{\alpha\beta}^{\mu\nu
},\left.  {}\right.  \left\{  g_{\alpha\beta},g_{\mu\nu}\right\}  =\left\{
p^{\alpha\beta},p^{\mu\nu}\right\}  =0. \label{eqnE20}%
\end{equation}

Note that the conjugate momenta have to be introduced for all generalized
coordinates irrespective of whether or not the corresponding time derivatives
(\textquotedblleft velocities\textquotedblright) for particular fields are
present in the Lagrangian. In fact, in the first order, metric-affine,
formulations of GR due to Einstein \cite{Einstein} (for English translation
see \cite{Einstein-eng}) some fields enter the Lagrangian with no derivatives,
but nevertheless momenta, conjugate to all of fields, are needed \cite{KK,
affine-metric}.

For regular, i.e. non-singular, systems, the two sets of phase-space variables
are related to each other by a canonical transformation if and only if the
following relations are fulfilled
\[
\left\{  g_{\alpha\beta},p^{\mu\nu}\right\}  _{g,p}=\left\{  g_{\alpha\beta
},p^{\mu\nu}\left(  \pi,g\right)  \right\}  _{g,\pi}=\Delta_{\alpha\beta}%
^{\mu\nu}~,
\]%
\begin{equation}
\left\{  g_{\alpha\beta},g_{\mu\nu}\right\}  _{g,p}=\left\{  g_{\alpha\beta
},g_{\mu\nu}\right\}  _{g,\pi}~=0,\left.  {}\right.  \left\{  p^{\alpha\beta
},p^{\mu\nu}\right\}  _{g,p}=\left\{  p^{\alpha\beta},p^{\mu\nu}\right\}
_{g,\pi}~=0. \label{eqnE21}%
\end{equation}
In our case, which is based on the use of phase-space variables of
\cite{Pirani} and \cite{Dirac}, we have to explicitly check in detail only one
PB (the rest of PBs is obviously fulfilled) to find
\begin{equation}
\left\{  p^{\gamma\sigma},p^{\delta\rho}\right\}  _{g,p}=\left\{  \pi
^{\gamma\sigma}-\frac{1}{2}\sqrt{-g}A^{\left(  \gamma\sigma\right)  0\mu\nu
k}g_{\mu\nu,k}~,~\pi^{\delta\rho}-\frac{1}{2}\sqrt{-g}A^{\left(  \delta
\rho\right)  0\alpha\beta m}g_{\alpha\beta,m}\right\}  _{g,\pi}=0.
\label{eqnE22}%
\end{equation}
Note that for pairs with at least one temporal index this PB was calculated in
\cite{Pirani, KKRV}, where it is just the PB between primary constraints%

\begin{equation}
\left\{  \phi^{\sigma0},\phi^{\gamma0}\right\}  =0. \label{eqn-star}%
\end{equation}
However, this result is also valid for all indices. The PB of (\ref{eqnE22})
shows that Dirac's modification of $L_{PSS}$ at the Lagrangian level leads to
a Hamiltonian formulation in which the phase-space variables are canonically
related to those of PSS.

Our next goal is to consider the effect of such a canonical change of
phase-space variables on all steps of the Dirac procedure. We can utilize the
PB of (\ref{eqnE22}), and, by rearranging terms, present the canonical part of
the PSS Hamiltonian in a different, but equivalent form by explicitly creating
(extracting) combinations that correspond to a canonical change of variables
\begin{equation}
\phi^{pq}=\pi^{pq}-\frac{1}{2}\sqrt{-g}A^{\left(  pq\right)  0\mu\nu k}%
g_{\mu\nu,k}~. \label{eqnE26}%
\end{equation}

This simple rearrangement is very convenient to study canonical
transformations and allows us to present two Hamiltonians of \cite{Pirani} and
\cite{Dirac} as one expression that will make transparent the effect of such
changes on all steps of the Dirac procedure. By substituting Eq.
(\ref{eqnE26}) into the canonical part of the PSS Hamiltonian, (\ref{eqn7}),
and using (\ref{eqnE11}) we obtain the total Hamiltonian, (\ref{eqn6-2}),
where $H_{c}$ written in terms of $\phi^{pq}$ takes the form
\[
H_{c}=\frac{1}{\sqrt{-g}g^{00}}I_{mnpq}\phi^{mn}\phi^{pq}-\frac{1}{g^{00}}%
\phi^{mn}\left(  g^{0t}g_{mn,t}-2g^{0\alpha}g_{\alpha n,m}\right)
\]%
\begin{equation}
+\frac{1}{4}\sqrt{-g}\left[  \frac{1}{g^{00}}\left(  g^{0k}E^{\left(
mn\right)  \mu\nu}-2g^{0\mu}E^{\left(  mn\right)  k\nu}\right)  \left(
g^{0t}\delta_{m}^{\alpha}\delta_{n}^{\beta}-2g^{0\alpha}\delta_{m}^{t}%
\delta_{n}^{\beta}\right)  -B^{\mu\nu k\alpha\beta t}\right]  g_{\mu\nu
,k}g_{\alpha\beta,t}~. \label{eqnE28}%
\end{equation}

Note that (\ref{eqn6-2}) and (\ref{eqnE28}) simultaneously represent the total
Hamiltonians for both formulations, \cite{Pirani} and \cite{Dirac}. In the
Dirac case $\phi^{\alpha\beta}=p^{\alpha\beta}$; while for PSS, $\phi
^{\alpha\beta}$ is given by (\ref{eqnE26}). Both equations (\ref{eqn6-2}) and
(\ref{eqnE28}) manifestly demonstrate the effect of canonical transformations
for the total Hamiltonians:
\begin{equation}
H_{T}^{PSS}\left(  g,\pi\right)  \mid_{g_{\mu\nu}=g_{\mu\nu};~p^{\gamma\sigma
}=\pi^{\gamma\sigma}-\frac{1}{2}\sqrt{-g}A^{\left(  \gamma\sigma\right)
0\mu\nu k}g_{\mu\nu,k}}=H_{T}^{D}\left(  g,p\right)  ; \label{eqnE29}%
\end{equation}
and for the generalized Hamiltonian equations:
\begin{equation}
g_{\alpha\beta,0}=\left\{  g_{\alpha\beta},H_{T}^{PSS}\right\}  ,\left.
{}\right.  \pi_{,0}^{\alpha\beta}=\left\{  \pi^{\alpha\beta},H_{T}%
^{PSS}\right\}  \Longrightarrow g_{\alpha\beta,0}=\left\{  g_{\alpha\beta
},H_{T}^{D}\right\}  ,\left.  {}\right.  p_{,0}^{\alpha\beta}=\left\{
p^{\alpha\beta},H_{T}^{D}\right\}  . \label{eqnE30}%
\end{equation}

In \cite{KKRV} the PSS formulation was analyzed by considering the
combinations of different orders in $\pi^{\alpha\beta}$. Here we will work in
orders of $\phi^{\alpha\beta}$ which, due to the simple relation $\left\{
\phi^{\alpha\beta}\left(  g,\pi\right)  ,\phi^{\mu\nu}\left(  g,\pi\right)
\right\}  _{\left(  g,\pi\right)  }=0$, makes the amount of calculations
absolutely the same for the PSS and Dirac formulations. It also makes
transparent the effect of the considered canonical transformation.

Now we calculate the time development of the primary constraints,
\[
\phi_{,0}^{0\sigma}=\left\{  \phi^{0\sigma},H_{c}\right\}  .
\]
By working with combinations $\phi^{\alpha\beta}$, we can use associative
properties of PB, and therefore $\left\{  \phi^{0\sigma},H_{c}\right\}
=-\frac{\delta H_{c}}{\delta g_{0\sigma}}$, where the variation is not
performed inside the expression for $\phi^{\alpha\beta}$, since $\left\{
\phi^{\alpha\beta},\phi^{\mu\nu}\right\}  =0$. The variation $-\frac{\delta
H_{c}}{\delta g_{0\sigma}}$ leads to the following secondary constraint
\[
\chi^{0\sigma}=-\frac{1}{2}\frac{1}{\sqrt{-g}}\frac{g^{0\sigma}}{g^{00}%
}I_{mnpq}\phi^{mn}\phi^{pq}+\delta_{m}^{\sigma}\left[  \phi_{,k}^{mk}+\left(
\phi^{pk}e^{qm}-\frac{1}{2}\phi^{pq}e^{km}\right)  g_{pq,k}\right]
\]%
\begin{equation}
+\frac{1}{2}\sqrt{-g}g^{0\sigma}\left[  -g_{mn,kt}E^{mnkt}+\frac{1}{4}%
g_{mn,k}g_{pq,t}\left(  -E^{mnpq}e^{kt}+2E^{ktpn}e^{mq}+E^{pqnt}e^{mk}\right)
\right]  . \label{eqnE32}%
\end{equation}
This expression coincides with the secondary constraint in Dirac's formulation
\cite{myths} where $\phi^{mn}=p^{mn}$. In order to show the equivalence of
(\ref{eqnE32}) to the secondary constraint of PSS \cite{KKRV}, one has to
rewrite this result in terms of $\pi^{km}$ using (\ref{eqnE26}).

Let us continue the Dirac procedure and consider the time development of the
secondary constraint
\begin{equation}
\chi_{,0}^{0\sigma}=\left\{  \chi^{0\sigma},H_{c}\right\}  +\left\{
\chi^{0\sigma},g_{00,0}\phi^{00}+2g_{0k,0}\phi^{0k}\right\}  . \label{eqnE33}%
\end{equation}
We have found it more convenient to perform the calculations in different
orders of momenta $\phi^{\mu\nu}$, which are indicated by the numbers in
brackets. We start from the PB of $\chi^{0\sigma}$with the primary constraint
for which the highest order contribution gives
\begin{equation}
\left\{  \chi^{0\sigma}\left(  2\right)  ,\phi^{0\gamma}\right\}  =-\frac
{1}{2}\frac{\delta}{\delta g_{0\gamma}}\left(  \frac{g^{0\sigma}}{\sqrt
{-g}g^{00}}\right)  I_{mnpq}\phi^{mn}\phi^{pq}=-\frac{1}{2}g^{\sigma\gamma
}\chi^{00}\left(  2\right)  . \label{eqnE34}%
\end{equation}

Using this higher order result, (\ref{eqnE34}), as a guide, we have to verify
by calculation that to all orders of $\phi^{km}$ this structure is preserved.
The explicit calculation confirms that the following PB is valid to all orders
of $\phi^{km}$
\begin{equation}
\left\{  \chi^{0\sigma},\phi^{0\gamma}\right\}  =-\frac{1}{2}g^{\gamma\sigma
}\chi^{00}. \label{eqnE35}%
\end{equation}
We obtained this relation for both formulations, \cite{KKRV} and \cite{myths},
and it demonstrates the form-invariance of the PB among the primary and
secondary constraints for canonically related formulations. Now we proceed and
find the PB of the secondary constraints with the canonical part of the
Hamiltonian $\left\{  \chi^{0\sigma},H_{c}\right\}  $. As before, we start
from the highest order contribution, which for this part is of third order in
$\phi^{ab}$,
\begin{equation}
\left\{  \chi^{0\sigma},H_{c}\right\}  \left(  3\right)  =\left\{
\chi^{0\sigma}\left(  2\right)  ,H_{c}\left(  2\right)  \right\}  =-\frac
{2}{\sqrt{-g}g^{00}}g^{\sigma d}\phi^{ab}I_{abcd}\frac{g^{0c}}{g^{00}}\left(
-\frac{1}{2}\frac{1}{\sqrt{-g}}I_{mnpq}\phi^{mn}\phi^{pq}\right)  ;
\label{eqnE36}%
\end{equation}
it can be presented as a term proportional to $\chi^{00}\left(  2\right)  $ or
$\chi^{0c}\left(  2\right)  $, or as a linear combination of both. So, we have
many possible and non-unique ways to present this result, which requires us to
investigate all combinations, to all orders. Such an approach involves a
considerable amount of calculation. The wrong choice can lead to the erroneous
conclusion that the time development of the secondary constraint
$\chi^{0\sigma}$ is not proportional to the secondary constraints and gives
rise to tertiary constraints, etc. The approach that allows one to perform
unambiguous calculations (sort out the contributions uniquely in terms of
secondary constraints) is presented in the Appendix and here we give only the
final result:
\[
\left\{  \chi^{0\sigma},H_{c}\right\}  =-\left[  \frac{2}{\sqrt{-g}}%
I_{pqmk}\frac{g^{\sigma m}}{g^{00}}\phi^{pq}+g^{0\sigma}g_{00,k}+2g^{\sigma
p}g_{0p,k}+g^{\sigma p}\frac{g^{0q}}{g^{00}}\left(  g_{pq,k}+g_{qk,p}%
-g_{pk,q}\right)  \right]  \chi^{0k}%
\]%
\begin{equation}
-\delta_{0}^{\sigma}\chi_{,k}^{0k}+\frac{1}{2}g^{\sigma k}g_{00,k}\chi^{00},
\label{eqnE37}%
\end{equation}
which is easy to compare with the results obtained for the Dirac and PSS
formulations in \cite{KKRV} and \cite{myths}. Note that again the constraints
and structure functions are different for the two formulations; but the whole
structure of (\ref{eqnE37}) is form-invariant. In fact, (\ref{eqnE37}) can be
presented in the following compact form $\left\{  \chi^{0\sigma}%
,H_{c}\right\}  =V_{\gamma}^{\sigma}\left(  g,\phi,\partial\right)
\chi^{0\gamma}$ where upon a canonical transformation not only the
constraints, but also the structure functionals $V_{\gamma}^{\sigma}$ of one
formulation transforms into another independently. Equation (\ref{eqnE37})
proves at the same time the closure of the Dirac procedure for both
formulations. This equation, along with (\ref{eqnE35}) and (\ref{eqn-star}),
is sufficient to find the gauge generators and derive the gauge
transformations for both formulations. We do not want to repeat such
calculations here, since they are given in detail in \cite{KKRV} and
\cite{myths} using two different methods described in \cite{Castellani} and
\cite{Novel}. The result of such calculations is the four-dimensional
diffeomorphism invariance (\ref{eqnE0}) (for both formulations) that follows
directly using each formalism, without any non-covariant and field-dependent
redefinition of the gauge parameters; and the gauge transformation can be
written in the covariant form (\ref{eqnE0}) for all components of the metric
tensor. For completeness we provide the expression for the canonical part of
the Hamiltonian
\begin{equation}
H_{c}=-2g_{0\sigma}\chi^{0\sigma}+\left(  2g_{0m}\phi^{mk}\right)
_{,k}-\left[  \sqrt{-g}E^{mnki}g_{mn,i}-\sqrt{-g}g_{\mu\nu,i}\frac{g^{0\mu}%
}{g^{00}}\left(  g^{\nu k}g^{0i}-g^{\nu i}g^{0k}\right)  \right]  _{,k}.
\label{eqnE38}%
\end{equation}
In both formulations, $H_{c}$ is the sum of the term proportional to the
secondary constraints, $-2g_{0\sigma}\chi^{0\sigma}$, and the total spatial
derivatives, despite having different expressions for $\chi^{0\sigma}$ and
$\phi^{mk}$ (for details see \cite{KKRV} and \cite{myths}).

\newpage

\section{Conclusion}

We have analyzed the relation between the two Hamiltonian formulations of GR,
\cite{Pirani} and \cite{Dirac}, which allow one to derive four-dimensional
diffeomorphism invariance \cite{KKRV, myths}. It has been shown that the full
sets of phase-space variables for these two formulations are related to each
other by a transformation of (\ref{eqnE18}), which satisfies the condition of
canonicity (\ref{eqnE21}) known for the Hamiltonian formulations of
non-singular systems. It also preserves the form-invariance of the expressions
for the total Hamiltonians (\ref{eqnE29}). These properties are well known for
Hamiltonian formulations of systems with regular (i.e., non-singular)
Lagrangians. Despite these similarities with regular systems, the analysis of
singular systems has a peculiarity; in the former, condition (\ref{eqnE21}) is
necessary and sufficient for equivalence of the two formulations, whereas in
the latter case (for singular systems) this condition is necessary, but not
sufficient. For singular systems it is also important to preserve the
form-invariance of the algebra of constraints, as is the case for the PSS and
Dirac formulations (see (\ref{eqn-star}), (\ref{eqnE35}), and (\ref{eqnE37})).
As we demonstrated in \cite{myths}, if one is tempted to convert the ADM
variables into canonical ones then, to satisfy (\ref{eqnE21}) all momenta have
to be involved, which leads to the relationships between \textquotedblleft
old\textquotedblright\ and \textquotedblleft new\textquotedblright\ momenta
given by Eqs. (163-165) of \cite{myths}%

\begin{equation}
p^{00}=-\Pi\frac{1}{2}\left(  -g^{00}\right)  ^{1/2}, \label{eqn450}%
\end{equation}

\begin{equation}
p^{0m}=\Pi\frac{1}{2}\left(  -g^{00}\right)  ^{-1/2}g^{0m}+\Pi_{k}\frac{1}%
{2}e^{km}, \label{eqn451}%
\end{equation}

\begin{equation}
p^{pq}=-\Pi\frac{1}{2}\left(  -g^{00}\right)  ^{-3/2}g^{0p}g^{0q}+\Pi_{k}%
\frac{1}{2}\left(  \frac{g^{0p}}{g^{00}}e^{kq}+\frac{g^{0q}}{g^{00}}%
e^{kp}\right)  +\Pi^{pq}, \label{eqn452}%
\end{equation}
or Eqs. (166-168) for the inverse transformations (see \cite{myths}, p. 608-609)%

\begin{equation}
\Pi=-2\left(  -g^{00}\right)  ^{-1/2}p^{00}, \label{eqn453}%
\end{equation}

\begin{equation}
\Pi_{n}=2g_{mn}p^{0m}+2g_{0n}p^{00}, \label{eqn454}%
\end{equation}

\begin{equation}
\Pi^{pq}=p^{pq}+\frac{g^{0q}}{g^{00}}\frac{g^{0p}}{g^{00}}p^{00}-\frac{g^{0p}%
}{g^{00}}p^{0q}-\frac{g^{0q}}{g^{00}}p^{0p}, \label{eqn455}%
\end{equation}
where $p^{\alpha\beta}$ and $\left\{  \Pi,\Pi_{k},\Pi^{pq}\right\}  $ are
momenta conjugate to $g_{\alpha\beta}$ and $\left\{  N,N^{k},g_{pq}\right\}
$, respectively.

This transformation (Eqs. (163-165) or Eqs. (166-168) of \cite{myths}) can be
called canonical, if only the PBs (\ref{eqnE21}) among phase-space variables
are concerned. However, as we have shown for the example of ADM gravity, even
if momenta conjugate to the ADM variables are introduced according to
(\ref{eqnE21}), which is a canonical transformation in the ordinary sense,
then these transformations nevertheless destroy the form-invariance of the
total Hamiltonian; the space-space \textquotedblleft
velocities\textquotedblright,\ which were eliminated at the first step of the
Dirac procedure, reappear again, and it is not clear what to do with them at
this stage of the Hamiltonian analysis (see Sec. 4.4 of \cite{myths} for the
detail discussion).

For non-singular systems, the canonicity condition (\ref{eqnE21}) is actually
independent of the particular form of the unconstrained Hamiltonian; and the
canonical transformations automatically convert the Hamiltonian, written in
terms of one set of phase-space variables, into another Hamiltonian. Whereas
for singular systems the Hamiltonian (that is, the total Hamiltonian) consists
of two distinct parts namely, the `canonical Hamiltonian' and a linear
combination of primary constraints, both of which play different roles. In
particular, the number of primary constraints corresponds to the number of
\textquotedblleft velocities\textquotedblright\ that cannot be solved in terms
of momenta, and, for systems with first-class constraints, it defines the
number of gauge parameters, an important intrinsic characteristic of a theory.
Thus, in the case of singular systems, the explicit form of the total
Hamiltonian becomes crucial and only transformations that preserve this form
(as in (\ref{eqnE29})) keep different formulations equivalent. This additional
condition makes the canonical transformations for singular systems much more
restrictive in comparison to non-singular systems.

We have considered two equivalent Hamiltonian formulations of GR, connected by
a relatively simple transformation that involves only a change of momenta. But
even from this simple case, it is possible to make a conjecture that the
equivalence of different Hamiltonian formulations of singular systems (at
least restricted to systems with only first-class constraints) is preserved if
the complete set of their phase-space variables is related by a canonical, in
ordinary sense, transformation, which in addition, must preserve the
form-invariance of the \textit{total} Hamiltonian and the algebra of
constraints. The transformation that does not satisfy these conditions will
not lead to a Hamiltonian formulation equivalent to the original theory
(Einstein's GR, as well as any other singular system).

Let us demonstrate our conjecture by the simple example, or the toy model
recently analyzed in \cite{ShestakovaCQG}, which illustrates the importance of
the compliance of both additional conditions: the preservation of the
form-invariance of the \textit{total} Hamiltonian and the algebra of
constraints. The following model is considered in \cite{ShestakovaCQG}:%

\begin{equation}
L_{1}=-\frac{1}{2}\frac{a\dot{a}^{2}}{N}+\frac{1}{2}Na~, \label{eqnE60}%
\end{equation}
which after the change of variables (\textquotedblleft
parametrization\textquotedblright, according to \cite{ShestakovaCQG})%

\begin{equation}
N=\sqrt{\mu}~,\text{ \ \ \ \ }a=a \label{eqnE61}%
\end{equation}
becomes%

\begin{equation}
L_{2}=-\frac{1}{2}\frac{a\dot{a}^{2}}{\sqrt{\mu}}+\frac{1}{2}\sqrt{\mu}a~.
\label{eqnE60a}%
\end{equation}

The Hamiltonian formulations for these two Lagrangians, (\ref{eqnE60}) and
(\ref{eqnE60a}), are presented in \cite{ShestakovaCQG} (Sec. 3) including the
restoration of gauge symmetries using the algorithm of \cite{Castellani}. The
two total Hamiltonians are:%

\begin{equation}
H_{T}^{\left(  1\right)  }=\dot{N}P-NT_{1}\text{ },\text{\ \ \ \ \ }%
H_{T}^{\left(  2\right)  }=\dot{\mu}\pi-2\mu T_{2}~, \label{eqnE62}%
\end{equation}
where $P$ and $\pi$ are momenta conjugate to $N$ and $\mu$, respectively; and
$T_{1}$ and $T_{2}$ are the secondary first class constraints of two
formulations (all details and explicit form of constraints can be found in
\cite{ShestakovaCQG}). The PBs among primary and secondary constraints for
these two formulations are:%

\begin{equation}
\left\{  P,T_{1}\right\}  =0\text{ },\text{\ \ \ \ }\left\{  \pi
,T_{2}\right\}  =\frac{1}{2\mu}T_{2}~. \label{eqnE63}%
\end{equation}

This difference in the algebra of constraints is responsible for distinct
transformations of these two formulations, because the generators are built
from all the first class constraints (this is Dirac's conjecture
\cite{Diracbook} realized as the algorithm in \cite{Castellani}).
Consequently, they depend on the algebra of PBs among all first class
constraints. From the canonicity condition, which reads for this model%

\begin{equation}
\dot{N}P=\frac{1}{2\sqrt{\mu}}\dot{\mu}P=\dot{\mu}\left(  \frac{1}{2\sqrt{\mu
}}P\right)  =\dot{\mu}\pi~, \label{eqnE64}%
\end{equation}
and using (\ref{eqnE61}) (in addition to (\ref{eqnE64})), we obtain the
relation between the momenta:%

\begin{equation}
P=2\sqrt{\mu}\pi. \label{eqnE65}%
\end{equation}
This phase-space transformation and its inverse leads to%

\begin{equation}
\left\{  N,P\right\}  _{\mu,\pi}=\left\{  \mu,\pi\right\}  _{N,P}%
=1.\label{eqnE66}%
\end{equation}
(Note that in this simple model, the transformations of $N$ and $P$ are
decoupled from the transformations of $a$ and $p$.)

The change of phase-space variables (\ref{eqnE61}) and (\ref{eqnE65}) can be
called canonical because of (\ref{eqnE66}), but this is not enough for there
to be equivalence of the two Hamiltonians (\ref{eqnE62}) because the algebra
of constraints (\ref{eqnE63}) is different, contrary to the preservation of
the constraint algebra for PSS and Dirac formulations (see the discussion
after (\ref{eqnE37})) that guarantees the equivalence of gauge
transformations. In \cite{ShestakovaGandC, ShestakovaCQG} great attention is
paid to the condition of canonicity (\ref{eqnE21}) and verification of it for
different parametrizations is emphasized. However, the example in
\cite{ShestakovaCQG} provides simple and explicit illustration of our point
that the notion of canonicity is more complicated for constrained Hamiltonian
systems and the ordinary condition on PBs is necessary, but not sufficient.
For this simple example the variables (\ref{eqnE61}) and (\ref{eqnE65}) have
canonical PBs (\ref{eqnE66}) and there is no need to use the effective
Lagrangian (Sec. 4 of \cite{ShestakovaCQG}) if only the PBs are concerned. The
canonical transformation between Dirac's variables (metric and the
corresponding momentum) and ADM is calculated in the extended phase-space
approach, and it was shown that it preserves the canonical structure of PBs.
However, we would like to mention that these transformations, Eqs. (12) and
(18) of \cite{ShestakovaGandC} and Eqs. (9) and (73) of \cite{ShestakovaCQG},
do not touch the variables from the gauge and ghost sectors of the extended
phase space and are in fact the same as we constructed before for ADM gravity
(see Eqs. (166-168) of \cite{myths}) without reference to any extension of a
phase space. In \cite{myths} this was done to illustrate that it is possible
to make changes from the Dirac to ADM phase-space variables in such a way that
the fundamental PBs preserve a canonical form, but this change destroys the
constraint structure and as such are not sufficient to ensure the equivalence
of the two formulations (see discussion on p. 609 of \cite{myths}). In the
extended phase-space approach, presented in \cite{ShestakovaGandC,
ShestakovaCQG}, this question cannot be analyzed because the Hamiltonian for
the effective action is presented neither in \cite{ShestakovaGandC}, nor in
\cite{ShestakovaCQG}. This was done only for the simple model described above,
where the PBs (\ref{eqnE66}) are canonical. The extended phase-space
Hamiltonian is written and the gauge transformations are obtained, but they
are different from the transformations that the author calls \textquotedblleft
correct\textquotedblright\ \cite{ShestakovaCQG}; and both are different from
the transformations that should be called \textquotedblleft
canonical\textquotedblright\ \cite{KKK}.

The motivation for the approach of \cite{ShestakovaGandC, ShestakovaCQG} was
based on the results of the ADM Hamiltonian formulation that is not related to
the Dirac \cite{Dirac, myths} and PSS \cite{Pirani, KKRV} formulations by a
canonical transformation \cite{myths, Fay, Italian}, and consequently, leads
to the gauge transformations which differ from diffeomorphism. An answer in
the affirmative to the question posed by Shestakova in \cite{ShestakovaCQG}%
\textit{\textquotedblleft Would not it be better to restrict ourself by
transformations in phase space of original canonical variables in the sense of
Dirac?\textquotedblright} (i.e. metric) cannot be reconciled with the
statement of \cite{ShestakovaCQG} \textquotedblleft the ADM parametrization is
more preferable\textquotedblright(but it does not lead to diffeomorphism when
the Dirac method is applied). The solution of this dilemma is proposed in
\cite{ShestakovaCQG}: \textquotedblleft we cannot consider the Dirac approach
as fundamental and undoubted\textquotedblright, and one ought to conclude
(based on this choice) that the ADM parametrization is, at least, more
fundamental and less doubted. So, as an alternative to the Dirac
parametrization-dependent method, another approach is proposed: the extended
phase space; by switching from the original Lagrangian to another, effective,
Lagrangian the equivalence of which to the original was not proven; further
many other questions, related to this proposal, remained unanswered in
\cite{ShestakovaGandC, ShestakovaCQG}, e.g.\textquotedblleft how to construct
a generator of gauge transformations in phase space\textquotedblright%
\ \cite{ShestakovaGandC}. It is difficult to discuss in detail the alternative
approach before, for example, the equivalence of the EH and effective action
is demonstrated. We argue that because the Dirac approach allows unique
selection of the canonical variables, it is more preferable than approaches
(if they exist) that lead to many canonical formulations. Our answer to the
above question (in italic) is \textquotedblleft YES\textquotedblright\ and,
because of this, we doubt that the approach proposed in \cite{ShestakovaGandC,
ShestakovaCQG} or any other approach \textquotedblleft will restore a
legitimate status of the ADM parametrization\textquotedblright%
\ \cite{ShestakovaCQG}. The detailed discussion of the many attempts to modify
the Dirac method to eliminate, what is, in our opinion, its the most
attractive feature, the uniqueness of a canonical formulation, is left for
another paper.

\section{Acknowledgment}

The authors are grateful to P.G. Komorowski, D.G.C. McKeon and A.V.
Zvelindovsky for helpful discussions. The partial support of The Huron
University College faculty of Arts and Social Science Research Grant Fund is
gratefully acknowledged.

\section{Appendix}

In this Appendix we describe the calculation of $\left\{  \chi^{0\sigma}%
,H_{c}\right\}  $, (\ref{eqnE37}), that proves the closure of the Dirac
procedure. This result is also needed to find generators and gauge
transformations which are discussed in detail for both formulations in
\cite{KKRV} and \cite{myths}.

As we have already pointed out, it is more convenient to perform the
calculations in different orders of momenta $\phi^{\mu\nu}$, which are
indicated by the numbers in brackets. Comparison of contributions of the
highest order in $\phi^{km}$ into the constraints $\chi^{0\sigma}$,
(\ref{eqnE32}), leads to a simple relation: $\chi^{0m}\left(  2\right)
-\frac{g^{0m}}{g^{00}}\chi^{00}\left(  2\right)  =0$. A generalization of this
result to all orders gives
\begin{equation}
\chi^{0m}-\frac{g^{0m}}{g^{00}}\chi^{00}=\psi^{0m} \label{eqnE39}%
\end{equation}
where
\begin{equation}
\psi^{0m}=\phi_{,k}^{mk}+\left(  \phi^{pk}e^{qm}-\frac{1}{2}\phi^{pq}%
e^{km}\right)  g_{pq,k}~. \label{eqnE40}%
\end{equation}

If the Dirac procedure is closed in terms of the constraints $\left(
\chi^{00},\chi^{0k}\right)  $, then it is also closed in terms of $\left(
\chi^{00},\psi^{0k}\right)  $, and vice versa, which follows directly from
(\ref{eqnE39}). However, working with the $\left(  \chi^{00},\psi^{0k}\right)
-$ pair allows us to sort terms unambiguously because we have only the
following non-zero contributions in both constraints: $\chi^{00}\left(
2\right)  $, $\chi^{00}\left(  0\right)  $, and $\psi^{0k}\left(  1\right)  $,
so separating terms of different order in $\phi^{km}$ simplifies calculations
(note that in \cite{KKRV} this procedure could not be simplified to such an extent).

Let us start with $\left\{  \chi^{00},H_{c}\right\}  $. In the highest order,
this PB, (\ref{eqnE36}), unambiguously gives (there are no derivatives of
$\phi^{km}$ in this expression, so it cannot be proportional to $\psi^{0k}$):
\begin{equation}
\left\{  \chi^{00},H_{c}\right\}  \left(  3\right)  =\left\{  \chi^{00}\left(
2\right)  ,H_{c}\left(  2\right)  \right\}  =-\frac{2}{\sqrt{-g}}\frac
{g^{0d}g^{0c}}{g^{00}g^{00}}\phi^{ab}I_{abcd}\chi^{00}\left(  2\right)  .
\label{eqnE41}%
\end{equation}
In the next order, we have contributions with and without derivatives of the
momenta $\phi^{km}$:
\begin{equation}
\left\{  \chi^{00},H_{c}\right\}  \left(  2\right)  =\left\{  \chi^{00}\left(
2\right)  ,H_{c}\left(  1\right)  \right\}  \left(  \phi\partial\phi\right)
+\left\{  \chi^{00}\left(  2\right)  ,H_{c}\left(  1\right)  \right\}  \left(
\phi\phi\right)  , \label{eqnE42}%
\end{equation}
which we consider separately starting from the terms proportional to
$\phi\partial\phi$. Such terms might be presented as proportional to the
corresponding orders of the $\left(  \chi^{00},\psi^{0k}\right)  -$ pair
through either derivatives of $\chi^{00}\left(  2\right)  $ or $\phi\phi
_{,k}^{mk}$, both of which have a particular structure in the indices. Again
this allows us to sort such terms uniquely:
\begin{equation}
\left\{  \chi^{00}\left(  2\right)  ,H_{c}\left(  1\right)  \right\}  \left(
\phi\partial\phi\right)  =-\frac{g^{0k}}{g^{00}}\chi_{,k}^{00}\left(
2\right)  \left(  \phi\partial\phi\right)  -\frac{2}{\sqrt{-g}}I_{mnpq}%
\phi^{mn}\frac{g^{0p}}{g^{00}}\psi^{0q}\left(  \partial\phi\right)
\label{eqnE43}%
\end{equation}
where
\begin{equation}
\chi_{,k}^{00}\left(  2\right)  =\chi_{,k}^{00}\left(  2\right)  \left(
\phi\partial\phi\right)  +\chi_{,k}^{00}\left(  2\right)  \left(  \phi
\phi\right)  =-\frac{1}{\sqrt{-g}}I_{mnpq}\phi^{mn}\phi_{,k}^{pq}-\frac{1}%
{2}\phi^{mn}\phi^{pq}\left(  \frac{1}{\sqrt{-g}}I_{mnpq}\right)  _{,k},
\label{eqnE50}%
\end{equation}%
\begin{equation}
\psi^{0m}=\psi^{0m}\left(  \partial\phi\right)  +\psi^{0m}\left(  \phi\right)
=\phi_{,k}^{mk}+\left(  \phi^{pk}e^{qm}-\frac{1}{2}\phi^{pq}e^{km}\right)
g_{pq,k}~. \label{eqnE51}%
\end{equation}

By performing the completion of (\ref{eqnE43}) to full expressions $\chi
_{,k}^{00}\left(  2\right)  $ and $\psi^{0q}\left(  1\right)  $, and combining
them with the second term of (\ref{eqnE42}), we obtain
\[
\left\{  \chi^{00}\left(  2\right)  ,H_{c}\left(  1\right)  \right\}  \left(
\phi\phi\right)  +\frac{g^{0k}}{g^{00}}\chi_{,k}^{00}\left(  2\right)  \left(
\phi\phi\right)  +\frac{2}{\sqrt{-g}}I_{mnpq}\phi^{mn}\frac{g^{0p}}{g^{00}%
}\psi^{0q}\left(  \phi\right)
\]
that, in case of closure, can be proportional only to $\chi^{00}\left(
2\right)  $. For the second order, we finally have
\[
\left\{  \chi^{00},H_{c}\right\}  \left(  2\right)  =-\frac{g^{0k}}{g^{00}%
}\chi_{,k}^{00}\left(  2\right)  -\frac{2}{\sqrt{-g}}I_{mnpq}\phi^{mn}%
\frac{g^{0p}}{g^{00}}\psi^{0q}%
\]%
\begin{equation}
+\left[  -\frac{g^{0k}g^{0\alpha}g^{0\beta}}{\left(  g^{00}\right)  ^{2}%
}g_{\alpha\beta,k}+\frac{1}{2}g^{0k}g_{00,k}-\left(  \frac{g^{0k}}{g^{00}%
}\right)  _{,k}\right]  \chi^{00}\left(  2\right)  . \label{eqnE52}%
\end{equation}
Next, the first order
\begin{equation}
\left\{  \chi^{00},H_{c}\right\}  \left(  1\right)  =\left\{  \chi^{00}\left(
2\right)  ,H_{c}\left(  0\right)  \right\}  +\left\{  \chi^{00}\left(
0\right)  ,H_{c}\left(  2\right)  \right\}  -\left(  Eq.(\ref{eqnE41})\right)
\left(  \chi^{00}\left(  2\right)  \rightarrow\chi^{00}\left(  0\right)
\right)  \label{eqnE53}%
\end{equation}
can be proportional to only $\psi^{0k}$, where the last term comes from the
completion of (\ref{eqnE41}) to the full (all orders) constraint $\chi^{00}$.
Here and in what follows the equations $\left(  Eq.(\#)\right)  $ inside the
formulae are used to indicate that the right-hand side of $\left(
Eq.(\#)\right)  $ must be substituted with the change indicated by
\textquotedblleft$\rightarrow$\textquotedblright.

In the last order, by using a similar compensation from the second order, we
have
\begin{equation}
\left\{  \chi^{00},H_{c}\right\}  \left(  0\right)  =\left\{  \chi^{00}\left(
0\right)  ,H_{c}\left(  1\right)  \right\}  -\left(  Eq.(\ref{eqnE52})\right)
\left(  \chi^{00}\left(  2\right)  \rightarrow\chi^{00}\left(  0\right)
,\psi^{0k}\rightarrow0\right)  \label{eqnE54}%
\end{equation}
which, in the case of closure, should give zero. This is confirmed by explicit calculation.

By calculating (\ref{eqnE53}) and combining it with the results of
(\ref{eqnE41}) and (\ref{eqnE52}), we obtain:
\[
\left\{  \chi^{00},H_{c}\right\}  =-\frac{2}{\sqrt{-g}}I_{mnpq}\phi^{pq}%
\frac{g^{0m}g^{0n}}{g^{00}g^{00}}\chi^{00}-\frac{g^{0k}}{g^{00}}\chi_{,k}%
^{00}-\frac{2}{\sqrt{-g}}I_{mnpq}\phi^{mn}\frac{g^{0p}}{g^{00}}\psi^{0q}%
\]%
\begin{equation}
+\left[  -\frac{g^{0k}g^{0\alpha}g^{0\beta}}{\left(  g^{00}\right)  ^{2}%
}g_{\alpha\beta,k}+\frac{1}{2}g^{0k}g_{00,k}-\left(  \frac{g^{0k}}{g^{00}%
}\right)  _{,k}\right]  \chi^{00}-\psi_{,k}^{0k}-\frac{g^{0\alpha}g^{0\beta}%
}{g^{00}}g_{\alpha\beta,t}\psi^{0t}, \label{eqnE55}%
\end{equation}
and in terms of $\chi^{00}$ and $\chi^{0k}$ equation (\ref{eqnE55}) gives:%

\begin{equation}
\left\{  \chi^{00},H_{c}\right\}  =-\left(  \frac{2}{\sqrt{-g}}I_{mnpk}%
\phi^{mn}\frac{g^{0p}}{g^{00}}+\frac{g^{0\alpha}g^{0\beta}}{g^{00}}%
g_{\alpha\beta,k}\right)  \chi^{0k}-\chi_{,k}^{0k}+\frac{1}{2}g^{0k}%
g_{00,k}\chi^{00}. \label{eqnE55a}%
\end{equation}

In a similar way, order by order, we consider the PB of $\psi^{0k}$ with
$H_{c}$, which leads to
\[
\left\{  \psi^{0m},H_{c}\right\}  =-\frac{2}{\sqrt{-g}}I_{pqkn}\phi^{pq}%
e^{mn}\psi^{0k}-\frac{g^{0m}}{g^{00}}\frac{g^{0k}}{g^{00}}\chi_{,k}^{00}%
+\frac{g^{0m}}{g^{00}}\psi_{,k}^{0k}+2\left(  \frac{g^{0k}}{g^{00}}\right)
_{,k}\psi^{0k}%
\]%
\begin{equation}
-\frac{g^{0p}}{g^{00}}e^{qm}\left(  g_{pk,q}-g_{pq,k}-g_{qk,p}\right)
\psi^{0k} \label{eqnE56}%
\end{equation}

\[
+\left[  \frac{g^{0m}}{g^{00}}\frac{g^{0q}}{g^{00}}e^{qk}g_{pq,k}+\frac
{1}{g^{00}}\left(  g^{0q}e^{km}-g^{0k}e^{qm}-g^{0m}e^{qk}\right)
g_{0q,k}+\frac{1}{2}e^{km}g_{00,k}\right]  \chi^{00}.
\]

Now, using Eq. (\ref{eqnE39}) and Eqs. (\ref{eqnE55}) - (\ref{eqnE56}) we find
$\left\{  \chi^{0m},H_{c}\right\}  $ in terms of $\chi^{00}$ and $\chi^{0k}$
\[
\left\{  \chi^{0m},H_{c}\right\}  =\left\{  \frac{g^{0m}}{g^{00}}\chi
^{00}+\psi^{0m},H_{c}\right\}  =\frac{g^{0m}}{g^{00}}\left\{  \chi^{00}%
,H_{c}\right\}  +\chi^{00}\left\{  \frac{g^{0m}}{g^{00}},H_{c}\right\}
+\left\{  \psi^{0m},H_{c}\right\}
\]%
\begin{equation}
=-\left[  \frac{2}{\sqrt{-g}}I_{pqkn}\phi^{pq}g^{mn}+g^{0m}g_{00,k}%
+2g^{pm}g_{0p,k}+\frac{g^{0p}}{g^{00}}g^{qm}\left(  g_{pk,q}+g_{pq,k}%
-g_{qk,p}\right)  \right]  \chi^{0k}+\frac{1}{2}g^{km}g_{00,k}\chi^{00}.
\label{eqnE57}%
\end{equation}
Finally, Eqs. (\ref{eqnE55a}) and (\ref{eqnE57}) which are written in terms of
$\left(  \chi^{00},\chi^{0k}\right)  $ can be combined into one covariant
expression expression, Eq. (\ref{eqnE37}), given in the main text.

\end{document}